\documentclass[sigconf]{acmart}
\AtBeginDocument{%
  }

\usepackage{multirow} 
\usepackage{graphicx}
\usepackage{subcaption}
\usepackage{enumitem}

\setcopyright{acmlicensed}
\copyrightyear{2018}
\acmYear{2018}
\acmDOI{XXXXXXX.XXXXXXX}
\acmConference[Conference acronym 'XX]{Make sure to enter the correct
  conference title from your rights confirmation email}{June 03--05,
  2018}{Woodstock, NY}
\acmISBN{978-1-4503-XXXX-X/2018/06}

\begin{document}

\title{ChoirRec: Semantic User Grouping via LLMs for Conversion Rate Prediction of Low-Activity Users}



\author{Dakai Zhai}
\authornote{Equal contributions.}
\affiliation{
  \institution{Alibaba Group}
  \city{Hangzhou}
  \country{China}
}
\email{zhaidakai.zdk@alibaba-inc.com}

\author{Jiong Gao}
\authornotemark[1]
\affiliation{
  \institution{Alibaba Group}
  \city{Hangzhou}
  \country{China}
}
\email{jionggao.gg@alibaba-inc.com}

\author{Boya Du}
\authornotemark[1]
\affiliation{
  \institution{Alibaba Group}
  \city{Hangzhou}
  \country{China}
}
\email{boya.dby@alibaba-inc.com}

\author{Junwei Xu}
\affiliation{
  \institution{SIGS, Tsinghua University}
  \city{Shenzhen}
  \country{China}
}
\email{xjw23@mails.tsinghua.edu.cn}

\author{Qijie Shen}
\affiliation{
  \institution{Alibaba Group}
  \city{Hangzhou}
  \country{China}
}
\email{qijie.sqj@alibaba-inc.com}

\author{Jialin Zhu}
\affiliation{
  \institution{Alibaba Group}
  \city{Hangzhou}
  \country{China}
}
\email{xiafei.zjl@alibaba-inc.com}

\author{Yuning Jiang}
\authornote{Corresponding author.}
\affiliation{
  \institution{Alibaba Group}
  \city{Hangzhou}
  \country{China}
}
\email{mengzhu.jyn@alibaba-inc.com}

\renewcommand{\shortauthors}{Zhai et al.}

\begin{abstract}
Accurately predicting conversion rates (CVR) for low-activity users remains a fundamental challenge in large-scale e-commerce recommender systems.
Existing approaches face three critical limitations: 
(i) reliance on noisy and unreliable behavioral signals; 
(ii) insufficient user-level information due to the lack of diverse interaction data; 
and (iii) a systemic training bias toward high-activity users that overshadows the needs of low-activity users.
To address these challenges, we propose ChoirRec, a novel framework that leverages the semantic capabilities of Large Language Models (LLMs) to construct semantic user groups and enhance CVR prediction for low-activity users.
With a dual-channel architecture designed for robust cross-user knowledge transfer, ChoirRec comprises three components: 
(i) a Semantic Group Generation module that utilizes LLMs to form reliable, cross-activity user clusters, thereby filtering out noisy signals; 
(ii) a Group-aware Hierarchical Representation module that enriches sparse user embeddings with informative group-level priors to mitigate data insufficiency; 
and (iii) a Group-aware Multi-granularity Modual that employs a dual-channel architecture and adaptive fusion mechanism to ensure effective learning and utilization of group knowledge. 
We conduct extensive offline and online experiments on Taobao, a leading industrial-scale e-commerce platform.
ChoirRec improves GAUC by 1.16\% in offline evaluations, while online A/B testing reveals a 7.24\% increase in order volume, highlighting its substantial practical value in real-world applications.
\end{abstract}




\maketitle

\section{Introduction}
\label{sec:intro}

\begin{figure}[t]
    \centering
    \begin{subfigure}[b]{0.49\columnwidth}
        \centering
        \includegraphics[width=\textwidth]{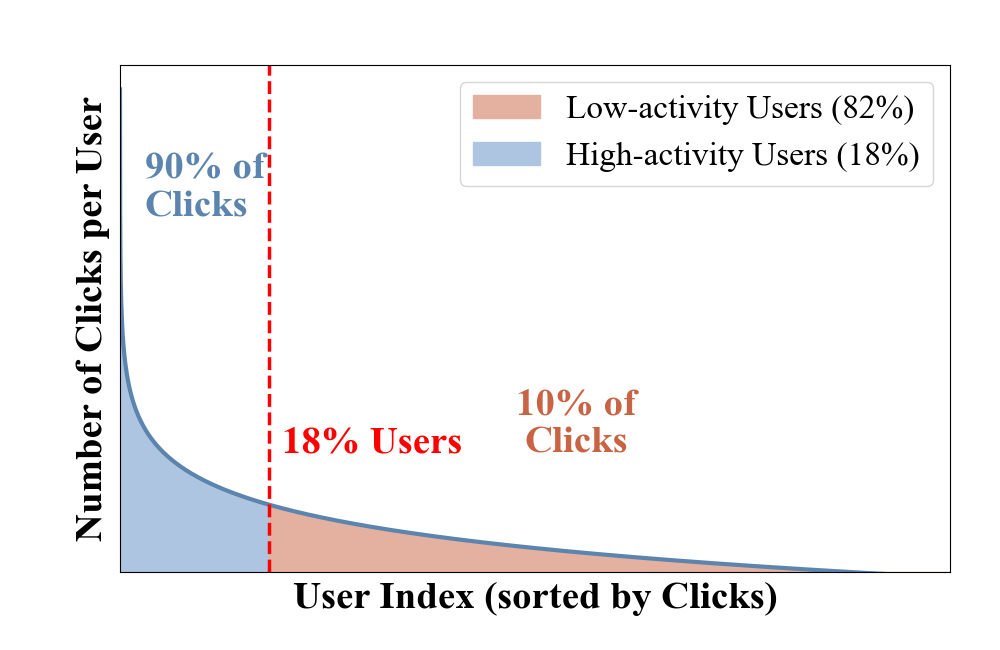}
        \caption{User clicks distribution}
        \label{fig:fenbu}
    \end{subfigure}
    \hfill
    \begin{subfigure}[b]{0.49\columnwidth}
        \centering
        \includegraphics[width=\textwidth]{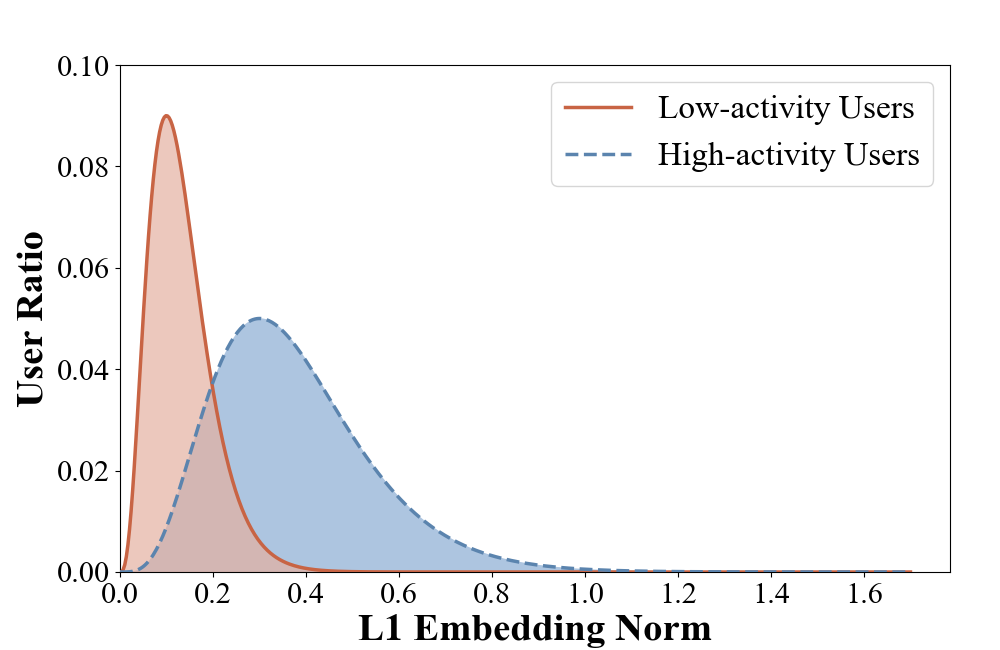}
        \caption{L1 norm distribution}
        \label{fig:fenbu2}
    \end{subfigure}
    
    \caption{User activity and embedding distributions}
    \label{fig:intro}
\end{figure}

Predicting conversion rates (CVR) for low-activity users is a fundamental challenge in e-commerce recommender systems. As illustrated in Figure~\ref{fig:fenbu}, user activity follows a severe long-tail distribution: the top 18\% of users account for 90\% of all clicks in the training data. This extreme skew results in severely under-represented and
unreliable user embeddings for the low-activity users.
This issue is empirically validated in Figure~\ref{fig:fenbu2}: The embeddings of low-activity users are smaller and more concentrated near zero. This indicates that their representations are severely under-trained and fail to capture rich preference signals, fundamentally undermining the reliability of CVR prediction.


While significant progress has been made in recommender systems, existing methods still face fundamental limitations when applied to low-activity users.
Knowledge transfer approaches~\cite{liu2025dynamic, li2019graph} aim to augment sparse user representations by leveraging collaborative signals from similar users or items.
However, their efficacy is inherently constrained by the noise and sparsity of the very data on which they rely. 
Meta-learning-based methods~\cite{lee2019melu} attempt to adapt model parameters to sparse contexts, yet they remain bottlenecked by the low information density of individual user histories. 
More recently, LLMs have shown promise in enriching user representations~\cite{liu2025llm-esr}, but current applications typically treat LLMs as vanilla feature extractors, failing to leverage their full semantic reasoning capabilities within a systematic framework. 
Consequently, the valuable signals derived from LLMs are often easily overshadowed.

We identify three challenges for effective CVR prediction on low-activity users:
(i) The knowledge sources are noisy and unreliable; 
(ii) User-level information is severely insufficient due to data sparsity; 
and (iii) Group-level and other signals enhancing low-activity users are overshadowed by individual signals during training. In Taobao’s production CVR model, for example, the GAUC for low‑activity users is over 5\% lower than that for highly active users, translating into a substantial business gap.

To overcome these challenges, we propose \textit{ChoirRec}, a novel LLM-driven framework that constructs high-quality semantic user groups to serve as a reliable knowledge source for long-tail user modeling. 
ChoirRec first leverages the semantic reasoning capacity of LLMs to generate expressive semantic profiles from raw behavioral and attribute data. 
These semantic profiles are then clustered using RQ-KMeans to form cross-activity semantic groups, thereby building a robust knowledge source and mitigating the impact of noisy signals.
Moreover, a \textit{group-aware hierarchical representation} module is adopted to directly enrich sparse individual user embeddings. 
It systematically generates multi-faceted group-level priors, including hierarchical group ID embeddings, aggregated semantic profiles, and group-level behavioral sequences.
Finally, a \textit{group-aware multi-granularity module} is designed to ensure effective integration of these group signals without being overshadowed.
This module employs a novel architecture featuring a dual-channel design.
Individual and group representations are learned in separate channels, with controllable and efficient knowledge transfer facilitated by an \textit{asymmetric information injection} mechanism and a \textit{gated knowledge distillation} strategy.

The main contributions of our work are summarized as follows:
\begin{itemize}
\item We propose \textit{ChoirRec}, a systematic framework that leverages LLMs to construct high-quality semantic user groups, establishing a reliable, high-quality knowledge source for low-activity users to alleviate both data sparsity and noise in CVR prediction.
\item We introduce a group-aware hierarchical representation module and a dual-channel modeling architecture. They jointly ensure well-formed semantic groups and effective leverage during training, preventing group signals from being dominated by individual-level noise.
\item We conduct extensive offline and online experiments on Taobao, a large-scale industrial e-commerce platform. ChoirRec achieves a 1.16\% improvement in GAUC for low-activity users in offline evaluation and drives a 7.24\% increase in order volume in online A/B tests, demonstrating its practical efficacy and significant business impact.
\end{itemize}

\section{Related Work}

Recommendation for low-activity users has attracted growing attention due to its practical importance and inherent challenges. 
Existing approaches can be broadly categorized into three lines of research: 
(1) knowledge transfer via data or representation enhancement, 
(2) specialized model architectures tailored for sparse data, 
and (3) recent efforts leveraging LLMs. 
Below, we review these directions and discuss their limitations, which motivate our proposed framework.

\subsection{Knowledge Transfer via Data and Representation Enhancement}

Knowledge transfer aims to improve the representation of data-scarce users by leveraging information from richer sources. 
In recommender systems, knowledge transfer is typically realized through either data augmentation or representation refinement.

Data augmentation methods enrich raw user inputs by synthesizing or rebalancing interactions. 
For example, Cold-Transformer~\cite{li2022transform} fuses positive and negative feedback to stabilize learning for cold users, while others generate pseudo-interactions to mitigate activity-level imbalances~\cite{chen2023www}.
As for the representation refinement, prior work enhances user embeddings by aggregating contextual or relational signals~\cite{yao2021selfsupervisedlearninglargescaleitem}, \emph{e.g.,} via clustering-based group profiling~\cite{liu2025dynamic}, Graph Neural Networks (GNNs) that capture high-order user-item dependencies~\cite{li2019graph}, or explicit knowledge distillation from high-activity to low-activity users~\cite{kim2023melt}. 

Despite their effectiveness, these methods remain fundamentally limited by the quality of the original data. 
When source data is sparse, noisy, or semantically shallow, knowledge transfer can suffer from \textit{negative transfer}, where misleading information degrades rather than improves performance. 
This highlights the necessity of a more robust and semantically meaningful knowledge source.

\subsection{Specialized Model Architectures for Data Sparsity}

Besides data-centric approaches, another line of work focuses on modeling architecture to better handle sparsity issues.~\cite{Zhang2021twotower, Wang2025kuai} 
Meta-learning frameworks~\cite{lee2019melu} treat each user as a few-shot learning task, enabling rapid adaptation from limited interactions. 
Other methods, like POSO~\cite{dai2021poso}, employ activity-level-specific subnetworks to explicitly separate learning for cold-start and other users, preventing the latter from dominating the training process.

While these methods effectively optimize the learning process, they largely provide a passive adaptation to sparsity but do not actively enrich input semantics. 
Under extreme sparsity or semantic ambiguity, even current advanced architectures struggle to recover reliable user interests. 
These underscore that refining model architectures alone is insufficient without simultaneously enhancing the semantic richness and reliability of user representations.

\subsection{LLM-Enhanced Recommendation}

The emergence of LLMs has introduced new opportunities to address data sparsity through their powerful semantic reasoning and generative capabilities. Current LLM-based recommendation methods fall into two categories:  

\begin{enumerate}
    \item \textbf{LLM-based data augmentation}, where LLMs generate synthetic user-item interactions or behavioral sequences~\cite{sun2025llmser, Wei2024LLMRec, liu2025filterllm, Gu2025R4ec}. 
While promising, such approaches often suffer from high inference costs, limited controllability, and potential hallucinations.

\item \textbf{LLM-based representation enhancement}, which uses LLMs to generate semantic embeddings for items with offline pipelines~\cite{liu2025llm-esr, liu2025llmalignmentlivestreamingrecommendation}. 
These representations are then integrated with collaborative signals in downstream models.
\end{enumerate}

However, most existing works treat LLMs as external feature extractors, integrating their outputs as static side information, rather than embedding their reasoning capabilities into downstream recommenders.
As a result, the full potential of LLMs to construct high-level semantic structures is underutilized. 
A principled framework that systematically leverages LLMs to build reliable group-level knowledge sources and seamlessly integrates them into CVR modeling remains an open challenge.

\section{Problem Formulation}
\label{sec:problem_formulation}

We consider a standard CVR prediction setting in e-commerce. 
Let $\mathcal{U}$ and $\mathcal{I}$ denote the sets of users and items, respectively. 
For a user--item pair $(u, i)$, the goal is to estimate the probability that user $u$ will purchase item $i$:

\begin{equation}
    \hat{y} = P(y=1 | u, i;\Theta),
\end{equation}
where $y \in \{0, 1\}$ indicates whether a purchase occurred.
\(\Theta\) is the model parameters.
Each user $u$ is associated with a static profile $P_u$ and a sequence of historical purchases $S_u^{\text{buy}}$.

We focus on the subset of \textit{low-activity users}, denoted as $\mathcal{U}_{\text{low}} \subset \mathcal{U}$, which is defined as users with significantly fewer interactions and sparser static attributes compared to the general population (\emph{e.g.}, those with only 19\% of the average purchase count and 65\% of attribute features). 
This data sparsity fundamentally limits the model's capacity to learn reliable preference representations, leading to suboptimal CVR prediction performance.

\section{Methodology}
\label{sec:methodology}

\begin{figure*}[t]
    \centering
    \includegraphics[width=\linewidth]{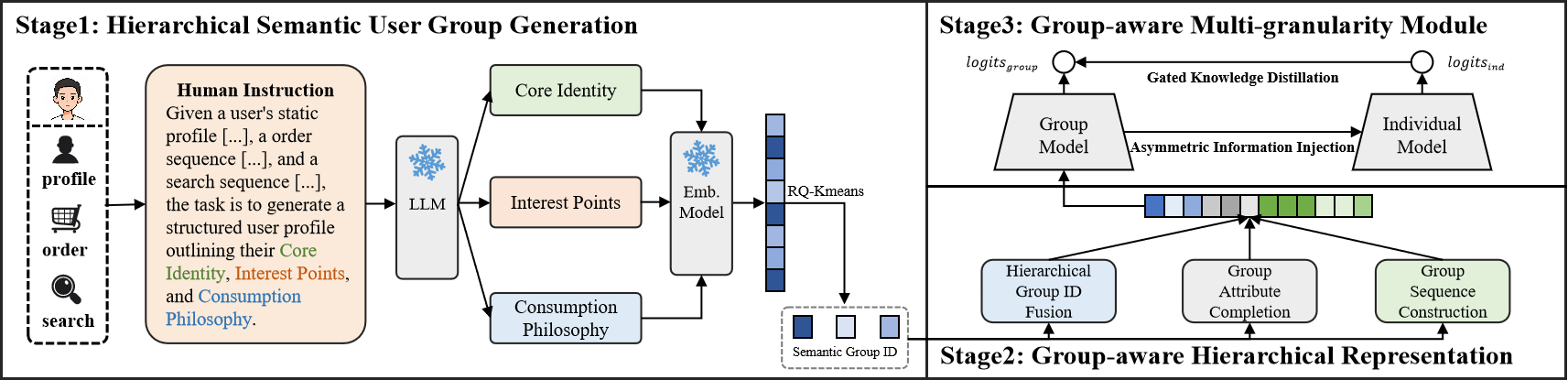}
    \caption{The overall architecture of the ChoirRec framework.}
    \label{fig:framework}
\end{figure*}

\subsection{Overview}
We now present \textbf{ChoirRec}, a novel framework for CVR prediction tailored to low-activity users. 
As illustrated in Figure~\ref{fig:framework}, ChoirRec follows a principled "generation--representation--modeling" pipeline that systematically addresses the three core challenges outlined in Section~\ref{sec:intro}: (i) noisy knowledge sources, (ii) insufficient user-level information, and (iii) overshadowing of group-level signals. 

ChoirRec constructs high-quality semantic user groups using LLMs and integrates them into a dual-channel architecture that ensures robust knowledge transfer. 
The framework comprises three modules:
(1) \textit{Semantic Group Generation} (Section~\ref{sec:group_generation}) leverages LLMs to form robust user groups, addressing the issue of noisy knowledge sources;
(2) \textit{Group-aware Hierarchical Representation} (Section~\ref{sec:representation}) enriches sparse user embeddings with group-level priors;
(3) \textit{Group-aware Multi-granularity Modeling} (Section~\ref{sec:modeling}) employs a decoupled dual-channel design with adaptive fusion to prevent group signals from being suppressed.

Below, we detail each component of our approach.

\subsection{Hierarchical Semantic User Group Generation}
\label{sec:group_generation}
To mitigate noisy and unreliable knowledge sources (Challenge i), the first stage constructs robust semantic user groups that establish a reliable basis for knowledge transfer.
A central distinction of our approach is the use of \emph{Semantic Groups} rather than \emph{Semantic IDs}.
Unlike Semantic IDs that serve as per-user learnable features, our Semantic Groups act as explicit units for knowledge aggregation and transfer, providing stable group-level priors to information-sparse users. 
To be effective, groups must exhibit strong generalization, robustness to noise, and cross-activity alignment. 
We achieve this with a two-stage \emph{synthesize--then--group} pipeline leveraging LLMs.

\subsubsection{\textbf{Semantic Profile Synthesis via LLM}}
We employ LLMs for this task because of their superior capabilities in reasoning, generalization, and noise suppression over heterogeneous inputs, whose capabilities are difficult to achieve with conventional statistical methods.

\noindentparagraph{\textbf{Input Formulation.}}
To effectively guide the LLM, we construct a textual input that contains three key components:
\begin{itemize}
    \item \textbf{Static Attributes:} The user's stable, non-behavioral information.
    \item \textbf{Time-windowed Aggregated Behaviors:} Historical purchases grouped by category and partitioned into recent, medium-term, and long-term windows to capture stable preferences while filtering transient noise.
    \item \textbf{Recent Search Queries:} Aggregated recent search queries that provide explicit, timely intent signals and mitigate sparsity in the user's purchase history.
\end{itemize}

\noindentparagraph{\textbf{Synthesis Process.}}
The LLM is instructed via a carefully designed prompt to generate a profile covering three dimensions: (i) core identity, (ii) key interest points, and (iii) consumption philosophy. 
The prompt encourages the LLM to leverage the world knowledge for generalization, explicitly ignoring outlier behaviors to denoise the input.
Applied uniformly across users, heterogeneous signals are projected into a shared semantic space, enabling consistent comparison between high- and low-activity users.

\subsubsection{\textbf{Hierarchical Group Construction}}
Following the profile generation, the second stage is to encode the profiles and conduct hierarchical grouping to obtain the final semantic user groups.

\begin{enumerate}
\item \textbf{Semantic Representation Encoding}: Each semantic profile $\mathcal{T}_u$ is encoded into a vector $e_u=\Phi_{\text{LLM}}(\mathcal{T}_u)\in \mathbb{R}^D$ using a powerful text embedding model (\emph{e.g.}, Qwen3-Embedding-8B).

\item \textbf{Hierarchical Grouping}:To form hierarchical semantic groups, we apply Residual Quantization K--Means (RQ--KMeans)~\cite{luo2024qarm}, decomposing high-dimensional clustering into $M$ sequential stages.
At stage $m$, the residual vector $r_u^{m-1}$ (with $r_u^{0}=e_u$) is clustered using KMeans, yielding a codeword index $\text{id}_u^m$ and updated residual $r_u^m$:
\end{enumerate}

\begin{align}
    \text{id}_u^m &= \underset{k}{\operatorname{argmin}} \| r_u^{m-1} - c_{m,k} \|^2, \\
    r_u^m &= r_u^{m-1} - c_{m,\text{id}_u^m}.
\end{align}

These steps generate the hierarchical group ID $G_u = (\text{id}_u^1, \dots, \text{id}_u^M)$, which serves as the index to a specific semantic group for downstream representation learning. RQ--KMeans is preferred over flat clustering as it produces a multi-granularity group ID—reducing misclassification of sparse low-activity users—and cuts the cost of clustering hundreds of millions of users in one step.

\subsection{Group-aware Hierarchical Representation}
\label{sec:representation}

With reliable semantic groups, we address insufficient user-level information (Challenge ii). 
Low-activity users exhibit sparsity in static attributes and behavioral histories, which hinders robust preference pattern learning. 
To this end, we construct a multi-faceted system of group-level priors derived from the semantic groups. 
These priors enrich sparse individual features via group aggregation, providing a stable and transferable knowledge source for the downstream model.

\begin{figure}[t]
    \centering
    \includegraphics[width=\columnwidth]{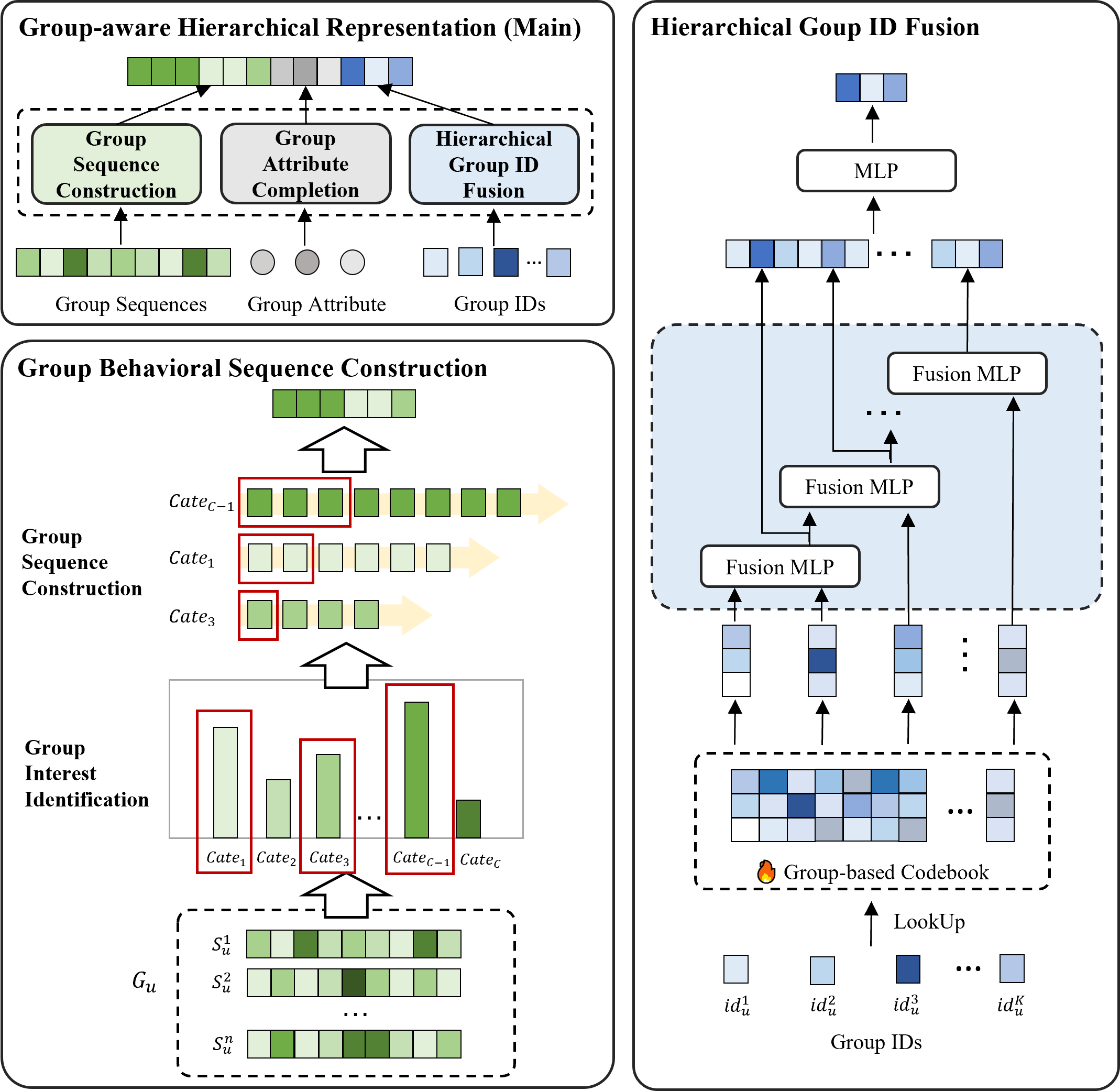}
    \caption{The construction of the Group-aware Hierarchical Representation system.}
    \label{fig:Sec3_3}
\end{figure}

As shown in Figure~\ref{fig:Sec3_3}, the feature system is constructed from the following three aspects, with each component addressing a distinct facet of information sparsity.

\subsubsection{\textbf{Hierarchical Group ID Fusion}}
We model a user's group ID $G_u$ as a discrete hierarchical code. 
Using level-specific embeddings independently fails to capture the hierarchical coarse-to-fine dependencies. 
To address this limitation, we introduce a prefix fusion mechanism to generate a hierarchy-aware group representation $\mathbf{e}_{G_u}$. 
The procedure is as follows:

\begin{enumerate}
    \item \textbf{Base Embedding}: For each user $u$, their group ID at level $l \in \{1, \dots, M\}$, denoted as $\text{id}_u^l$, is mapped to a base embedding vector by looking up a level-specific codebook $\mathbf{C}^{(l)} \in \mathbb{R}^{K \times d_e}$:
    \begin{equation}
        \mathbf{e}_{\text{base}}^{(l)} = \mathbf{E}^{(l)}[\text{id}_u^l]
    \end{equation}

    \item \textbf{Hierarchical Fusion}: A set of fused embedding vectors $\{\mathbf{e}_{\text{fuse}}^{(1)}, \dots, \mathbf{e}_{\text{fuse}}^{(M)}\}$ are then iteratively generated. 
    For any level $l>1$, the fused embedding $\mathbf{e}_{\text{fuse}}^{(l)}$ is generated by a fusion network $\mathcal{F}^{(l)}$ that combines the fused embedding of its parent level, $\mathbf{e}_{\text{fuse}}^{(l-1)}$, with the base embedding of the current level, $\mathbf{e}_{\text{base}}^{(l)}$:
    \begin{equation}
        \mathbf{e}_{\text{fuse}}^{(l)} = \tanh(\mathbf{W}^{(l)} [\mathbf{e}_{\text{fuse}}^{(l-1)} ; \mathbf{e}_{\text{base}}^{(l)}] + \mathbf{b}^{(l)})
    \end{equation}
    where $\mathbf{e}_{\text{fuse}}^{(1)} = \mathbf{e}_{\text{base}}^{(1)}$, $[\cdot ; \cdot]$ denotes concatenation, and $\mathbf{W}^{(l)}, \mathbf{b}^{(l)}$ are learnable parameters. This fusion ensures each level's representation includes context from all coarser parent levels.

    \item \textbf{Final Aggregation}: Finally, all the hierarchy-aware embeddings are concatenated and projected through a Multi-Layer Perceptron (MLP) to capture non-linear interactions across different granularities. 
    L2 normalization is applied to the final vector to stabilize training, yielding the group ID representation $\mathbf{e}_{G_u}$.
\end{enumerate}

\subsubsection{\textbf{Group Attribute Completion}}
 To address the insufficiency of individual static attribute features, we leverage the statistical properties of the group to construct a robust attribute prior. For each user $u$'s group $G_u$, we pre-calculate a stable group attribute feature $\mathbf{e}_{P_{G_u}}$ by aggregating the static attributes of all its members.
\begin{itemize}
    \item For discrete attribute fields, we use the mode as the group's representative value.
    \item For continuous attribute fields, we use the mean as the group's representative value.
\end{itemize}
Consequently, when an attribute field is missing for a low-activity user, the model leverages group-level statistics as a reliable substitute, thereby mitigating the challenges induced by sparsity in individual static attributes.

\subsubsection{\textbf{Group Behavioral Sequence Construction}}
The extreme sparsity of a low-activity user's purchase history is the primary bottleneck for interest modeling. To directly address this, we construct a dense behavioral sequence $S_{G_u}$ 
in two stages to represents the common interests of the group:

\begin{enumerate}
    \item \textbf{Group Interest Identification}: We first analyze the purchase history of all users within group $G_u$ and aggregate items by their categories to identify the Top-K most frequent interest categories for that group.
    \item \textbf{Group Sequence Construction}: From these $K$ high-frequency categories, the most popular items are then selected to form a representative group behavioral sequence. To integrate temporal nature into the sequence, we sort these items by their average purchase timestamp within the group.
\end{enumerate}
The constructed group sequence $S_{G_u}$ provides a dense, high-quality behavioral signal for low-activity users. In the downstream model, it serves as a robust complement to the user's sparse history, enabling the model to capture potential purchase intentions that would otherwise be undiscoverable.

\subsection{Group-aware Multi-granularity Module}
\label{sec:modeling}

\begin{figure}[t]
    \centering
    \includegraphics[width=1.00\columnwidth]{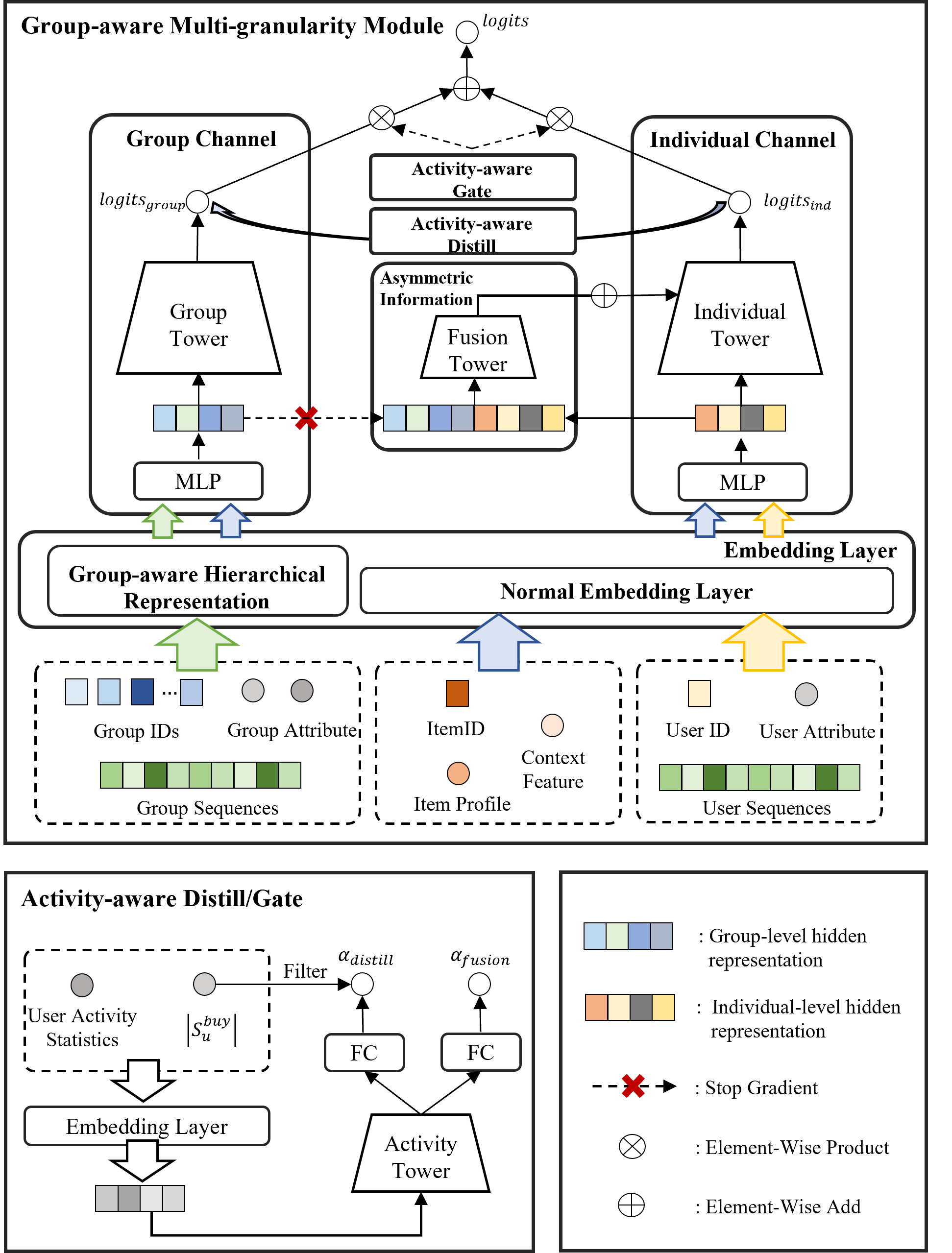}
    \caption{The architecture of the Group-aware Multi-granularity Module}
    \label{fig:Sec3_4}
\end{figure}

Having established robust group-level priors, the remaining challenge is to integrate them effectively into the training process without allowing their influence to be overshadowed (Challenge iii).
In joint training, gradients are often dominated by high-activity users. Their abundant and predictive individual-level signals naturally steer the model’s optimization toward individual features, leaving group-level knowledge underutilized. This bias is particularly detrimental to low-activity users, who rely on these stable group priors to compensate for sparse personal data.

To address this issue, we design a group-aware multi‑granularity module. As illustrated in Figure~\ref{fig:Sec3_4}, the module adopts a dual-channel structure: individual and group preferences are modeled in separate pathways, with carefully designed mechanisms to encourage selective and effective cross-channel interaction.



\subsubsection{\textbf{Dual-Channel Representation}}
The multi-granularity module employs a dual-channel architecture, consisting of an individual and a group channel that share a common item representation.

\noindentparagraph{\textbf{Individual Channel.}}The individual channel captures a user’s personal, dynamic preferences. It embeds the user’s static features (e.g., ID and attributes $P_u$) and contextual information into dense vectors, and applies a target‑attention mechanism~\cite{Zhou2017DIN} to the sparse purchase sequence $S_u^{\text{buy}}$ to produce a candidate-conditioned interest representation. The non‑sequential embeddings, target‑aware sequential vector, and candidate‑item features are concatenated to form the final individual representation $\mathbf{h}_u^{\text{ind}} \in \mathbb{R}^d$.

\noindentparagraph{\textbf{Group Channel.}} Complementing the individual pathway, the group channel focuses on the stable, shared preferences of the user’s semantic group. Its inputs are the fused group ID embedding $\mathbf{e}_{G_u}$, aggregated attribute vector $\mathbf{e}_{P_{G_u}}$, group-level behavioral sequence $S_{G_u}$) and candidate‑item features.  Following the same process as the individual channel, it produces non‑sequential and sequence representations, combined into the group‑level embedding $\mathbf{h}_u^{\text{group}} \in \mathbb{R}^d$.

\subsubsection{\textbf{Asymmetric Information Injection}}
To enable the individual channel to leverage stable group priors, we introduce the \textit{asymmetric information injection} mechanism. This creates a one‑way flow of information from the group channel to the individual channel: the latter is augmented with robust group knowledge, while the former remains isolated to preserve stability.

Concretely, we extract intermediate-layer outputs from both the group channel ($\mathbf{h}_{u, \text{mid}}^{\text{group}}$) and the individual channel ($\mathbf{h}_{u, \text{mid}}^{\text{ind}}$). These are concatenated and processed by a two‑layer MLP to generate a fused representation:
\begin{equation}
\mathbf{h}_{u}^{\text{fuse}} = \mathbf{W}_2 \cdot \text{ReLU}(\mathbf{W}_1 [\mathbf{h}_{u, \text{mid}}^{\text{group}} ; \mathbf{h}_{u, \text{mid}}^{\text{ind}}] + \mathbf{b}_1) + \mathbf{b}_2
\end{equation}
where $\mathbf{W}_1, \mathbf{b}_1, \mathbf{W}_2$, $\mathbf{b}_2$ are learnable parameters. The fused vector is injected into a later layer of the individual channel, $\mathbf{h}_{u, \text{post}}^{\text{ind}}$, via an additive connection.

\subsubsection{\textbf{Gated Knowledge Distillation}}
To enhance the group channel (student), we adopt \textit{gated knowledge distillation}, where the individual channel (teacher) provides soft supervision only from reliable high‑activity cases.

\noindentparagraph{\textbf{Margin-based Distillation Loss.}}
Standard KL‑divergence is unsuitable in our setting, as the teacher and student consume different input sources, which can lead to unstable optimization and gradient explosion when their predictions differ greatly. Thus, we use a more stable \textit{margin‑based squared‑error} loss:
\begin{equation}
    \mathcal{L}_{\text{margin}} = \max\left(0, \left|\sigma\left(\frac{z_{\text{ind}}}{T}\right) - \sigma\left(\frac{z_{\text{group}}}{T}\right)\right| - m\right)^2
\end{equation}
where $T$ denotes the temperature and $m$ is a fixed margin. This formulation offers two key advantages: (i) the margin $m$ tolerates small discrepancies, preventing the student from over‑fitting to the teacher; and (ii) because the derivative of the sigmoid diminishes as predictions approach 0 or 1, extreme disagreements do not trigger large gradients, ensuring stable training.

\noindentparagraph{\textbf{Dual Gating Mechanism.}}
Distillation is further controlled by two gates:
\begin{itemize}
    \item \textbf{Qualification Gate ($g_{\text{qual}}$):} A hard gate that determines whether the teacher’s output is eligible for distillation. Qualification requires the user to be highly active ($|S_u^{\text{buy}}| \geq \theta_{\text{act}}$) and the teacher’s prediction to be confident ($|\sigma(z_{\text{ind}}) - 0.5| > \theta_{\text{conf}}$).
    \begin{equation}
        g_{\text{qual}} = \mathbb{I}(|S_u^{\text{buy}}| \geq \theta_{\text{act}} \land |\sigma(z_{\text{ind}}) - 0.5| > \theta_{\text{conf}})
    \end{equation}

    \item \textbf{Reliability Gate ($\alpha_{\text{distill}}$):} A soft gate generated by a reliability network conditioned on user‑activity features, assigning a dynamic weight in $[0,1]$ to control distillation strength.
\end{itemize}
The effective distillation loss is then given by:
\begin{equation}
\mathcal{L}_{\text{KD}} = g_{\text{qual}} \cdot \alpha_{\text{distill}} \cdot \mathcal{L}_{\text{margin}}.
\end{equation}
This dual‑gate design ensures that knowledge is distilled only when the teacher is reliable and that the intensity of supervision is adaptively tuned, safeguarding the group channel from noisy or misleading signals.

\subsubsection{\textbf{Final Prediction and Optimization}}
The final prediction is an adaptive fusion of the two channels's logits, with a soft gating weight $\alpha_{\text{fusion}}$ --- produced by a dedicated head within the same reliability network used for distillation --- controlling the contribution of each channel:
\begin{equation}
    z_{\text{fused}} = (1 - \alpha_{\text{fusion}}) \cdot z_{\text{ind}} + \alpha_{\text{fusion}} \cdot z_{\text{group}}
\end{equation}
The model thus relies more on the group channel for low‑activity users and more on the individual channel for high‑activity users.

Training minimizes
\begin{equation}
\mathcal{L} = \mathcal{L}_{\text{BCE}} + \lambda \cdot \mathcal{L}_{\text{KD}}
\end{equation}
where
\begin{equation}
\mathcal{L}_{\text{BCE}} = - \frac{1}{N} \sum_{i=1}^{N} [y_i \log(\hat{y}_i) + (1 - y_i) \log(1 - \hat{y}_i)]
\end{equation}
$\hat{y}_i = \sigma(z_{\text{fused}}^{(i)})$ denotes the predicted probability for sample $i$ given ground‑truth label $y_i$, and $\lambda$ serving as a tunable coefficient to balance the influence of the distillation term against the primary supervised‑learning loss. This design ensures that adaptive fusion and cross‑channel knowledge transfer are jointly optimized, yielding a model that can flexibly exploit either channel depending on user‑activity level.

\section{Experiments}
\label{sec:experiments}

In this section, we conduct extensive offline and online experiments to evaluate the effectiveness of ChoirRec.

\subsection{Experimental Setup}

\noindentparagraph{\textbf{Dataset.}}
The offline evaluation is performed on a 14‑day dataset collected from the Taobao platform’s click‑log records, containing tens of billions of user interactions. The first 13 days are used for model training and the final day is reserved for testing. Following our problem formulation, \textit{low‑activity users} are defined as those with sparse recent behaviors; They account for roughly 55\% of the overall user base and suffers from pronounced data sparsity.

\noindentparagraph{\textbf{Evaluation Metrics.}}
Model performance is assessed using two ranking‑quality metrics, with \textit{GAUC} as the primary measure:
\begin{itemize}
    \item \textbf{AUC (Area Under the ROC Curve):} A standard metric for overall ranking quality.
    \item \textbf{GAUC~\cite{Zhou2017DIN} (User-Weighted AUC):} Our primary metric. It computes a user‑weighted average of individual AUCs, providing a fairer evaluation of personalized ranking quality.

\end{itemize}

\noindentparagraph{\textbf{Baselines.}}
We compare \textit{ChoirRec} against a set of strong baselines, including the production model and recent methods specifically designed to address user‑activity sparsity:
\begin{itemize}
    \item \textbf{Base Model:} The online production model.
    \item \textbf{POSO~\cite{dai2021poso}:} Mitigates high‑activity user dominance via separate sub‑networks for different activity levels.
    \item \textbf{Cold-Transformer~\cite{li2022transform}:} A Transformer‑based model that enriches representations of sparse users by fusing multiple behavior types.
    \item \textbf{MELT~\cite{kim2023melt}}: Enhances long‑tail user/item representations using a dual‑branch knowledge‑transfer architecture.
    \item \textbf{UIE~\cite{Liu2024enhancing}}: Recovers missing user interests by retrieving representative cluster centroids from a memory network.
\end{itemize}

\noindentparagraph{\textbf{Implementation Details.}}
For semantic‑group generation, we employ Qwen3‑30B‑A3B~\cite{yang2025qwen3technicalreport} to synthesize the LLM‑based semantic profile for each user, and encode profiles with the embedding model Qwen3‑Embedding‑8B~\cite{zhang2025qwen3embeddingadvancingtext}. Leveraging the model’s \textit{Matryoshka} property~\cite{kusupati2022matryoshka}, we truncate the original 4096‑dimensional embeddings to 512 dimensions, maintaining semantic quality while improving efficiency. Hierarchical grouping is performed using RQ‑KMeans in three stages, each with 256 centroids.

The core components of \textit{ChoirRec}—including the individual channel, group channel, reliability network, and asymmetric injection network—are all implemented as multi‑layer MLPs.

For training, we use the Adagrad optimizer~\cite{John2011Adagrad} with a batch size of 1024. The learning rate starts at 0.01 and gradually decays to 0.001 over the course of training.

\subsection{Overall Performance}
\label{sec:main_results}

We compare \textit{ChoirRec} with several strong baselines, as summarized in Table~\ref{tab:main_performance}. Across both ranking metrics (\textit{AUC} and \textit{GAUC}) and for both low‑activity and overall users, \textit{ChoirRec} delivers consistent, statistically significant gains.

Relative to the Base Model, the GAUC improvement reaches 1.16\% for low‑activity users and 0.47\% overall, demonstrating the value of integrating group‑level semantic knowledge into CVR prediction. The framework also outperforms recent state‑of‑the‑art approaches across all measured metrics. The largest gains are observed in the low‑activity segment, indicating that semantic grouping is particularly effective in high‑sparsity scenarios—the core problem this work aims to solve.

\begin{table}[t]
\centering
\caption{Overall performance comparison. Best results are in \textbf{bold}; second-best are \underline{underlined}.}
\label{tab:main_performance}
\begin{tabular*}{\columnwidth}{@{\extracolsep{\fill}} lcccc}
\toprule
\multirow{2}{*}{\textbf{Model}} & \multicolumn{2}{c}{\textbf{AUC}} & \multicolumn{2}{c}{\textbf{GAUC}} \\
\cmidrule(lr){2-3} \cmidrule(lr){4-5}
& \textbf{Low-Act.} & \textbf{Overall} & \textbf{Low-Act.} & \textbf{Overall} \\
\midrule
Base Model & 0.9195 & 0.9098 & 0.7097 & 0.7732 \\
\addlinespace
POSO & 0.9199 & 0.9100 & 0.7111 & 0.7729 \\
Cold-Transformer & 0.9198 & 0.9099 & 0.7103 & 0.7734 \\
MELT & 0.9201 & \underline{0.9103} & 0.7119 & 0.7746 \\
UIE & \underline{0.9203} & 0.9102 & \underline{0.7132} & \underline{0.7750} \\
\addlinespace
\textbf{ChoirRec (Ours)} & \textbf{0.9225} & \textbf{0.9116} & \textbf{0.7179} & \textbf{0.7768} \\
\bottomrule
\end{tabular*}
\end{table}

\subsection{Performance across User Activity Levels}
\label{sec:activity_analysis}

To evaluate \textit{ChoirRec} under different levels of data sparsity, we divide test users into five equally sized groups based on historical purchase frequency. Level 1 represents the lowest‑activity users, while Level 5 contains the most active. For each Level, we measure the relative GAUC improvement of \textit{ChoirRec} over the Base Model.

\begin{figure}[t]
    \centering
    \includegraphics[width=0.95\columnwidth]{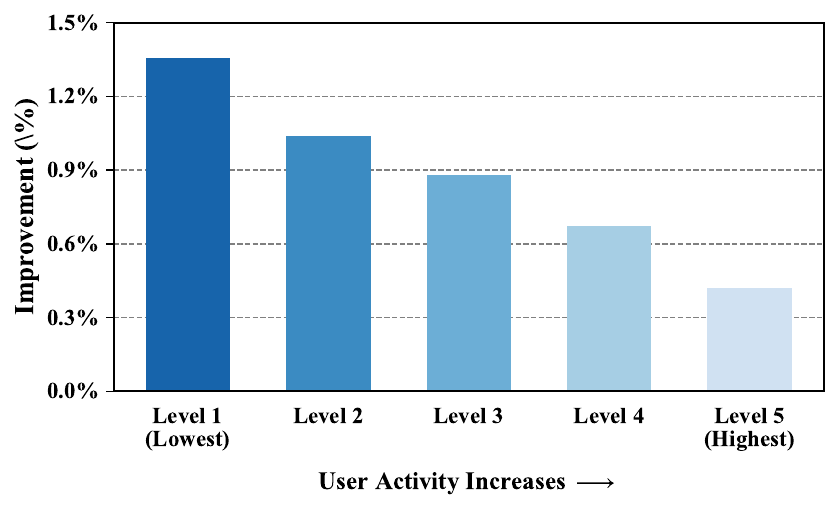}
    \caption{Relative GAUC improvement of ChoirRec over the Base Model across five user activity levels, ordered by historical purchase frequency (Level 1 = lowest activity, Level 5 = highest).}
    \label{fig:activity_analysis}
\end{figure}

As shown in Figure~\ref{fig:activity_analysis}, \textit{ChoirRec} , the model achieves gains across all Levels, with the largest improvement in the sparsest segment: GAUC rises by 1.355\% for Level 1 and gradually decreases to 0.420\% for Level 5. This inverse relationship between activity level and gain supports our design, showing that the group‑aware framework is particularly effective when individual behavioral signals are limited—precisely the condition it was built to address. The fact that gains remain positive even for the most active users further demonstrates that group‑level context adds value in data‑rich scenarios, underscoring the broad applicability of our approach.

\subsection{Ablation Study}
\label{sec:ablation_study}

We perform an ablation study to evaluate the contribution of each core component in \textit{ChoirRec}. The results in Table~\ref{tab:ablation_study} show that removing any single component leads to a clear drop in GAUC, confirming both its necessity and effectiveness.

The impact of each component group is detailed below:
\begin{itemize}
    \item \textbf{Semantic Group Generation:} In the variant \textit{w/o LLM Emb.}, the LLM‑generated semantic embeddings used for grouping are replaced with simple user‑ID embeddings.  This substitution causes a GAUC decline of 0.90\% for low‑activity users, underscoring the importance of the high‑quality semantic information synthesized by the LLM.
    
    \item \textbf{Group-aware Hierarchical Representation:} We remove each of the three group‑level priors individually. The largest drop occurs when the group behavioral sequence (\textit{w/o Seq. Emb.}) is omitted, as it directly supplements sparse user histories with dense behavioral signals. Excluding the group ID (\textit{w/o ID Emb.}) or the aggregated group attributes (\textit{w/o Attr. Emb.}) also produces measurable declines in GAUC, confirming the value of a multi‑faceted group representation system.  
    
    \item \textbf{Knowledge Integration:} Replacing our custom distillation loss with standard KL divergence (\textit{w/ KL Loss}) results in a 1.63\% GAUC drop for low‑activity users, showing that conventional losses are unstable in this scenario. Removing the dual‑channel architecture (\textit{w/o Dual‑Channel}) leads to significant decreases across all user levels, as the absence of a separate group channel causes the group‑level representations to be overshadowed by stronger individual‑level signals, reducing their contribution to the final prediction. Within the collaboration mechanisms, eliminating gated distillation primarily hurts low‑activity performance by weakening the group channel’s training, while removing asymmetric injection reduces gains for all levels by limiting the individual channel’s access to group‑level priors.  
\end{itemize}

\begin{table}[t]
\centering
\caption{Ablation study of ChoirRec. The table shows the relative GAUC change (\%) compared to the full model on both Low-activity and Overall user segments.}
\label{tab:ablation_study}
\begin{tabular*}{\columnwidth}{@{\extracolsep{\fill}} l cc}
\toprule
\multirow{2}{*}{\textbf{Model Variant}} & \multicolumn{2}{c}{\textbf{GAUC Change}} \\
\cmidrule(lr){2-3}
& \textbf{Low-Act.} & \textbf{Overall} \\
\midrule
\textbf{ChoirRec (Full Model)} & \textbf{0.00\%} & \textbf{0.00\%} \\
\addlinespace
\textit{Hierarchical Semantic Group Generation:} & & \\
\quad w/o LLM Emb. & -0.90\% & -0.36\% \\
\addlinespace
\textit{Group-aware Hierarchical Representation:} & & \\
\quad w/o ID Emb. & -0.24\% & -0.06\% \\
\quad w/o Attr. Emb. & -0.44\% & -0.09\% \\
\quad w/o Seq. Emb. & -0.59\% & -0.19\% \\
\addlinespace
\textit{Group-aware Multi-granularity Module:} & & \\
\quad w/o Dual-Channel & -0.67\% & -0.17\% \\
\quad w/o Gated Distillation & -0.43\% & -0.09\% \\
\quad w/o Asymmetric Injection & -0.25\% & -0.21\% \\
\quad w/ \: KL Loss & -1.63\% & -1.06\% \\
\quad w/o Margin & -0.21\% & -0.11\% \\
\bottomrule
\end{tabular*}
\end{table}

\subsection{Further Analysis}
\label{sec:analysis}

We provide additional analyses to better understand \textit{ChoirRec}’s behavior and verify the effectiveness of its key design choices.

\noindentparagraph{\textbf{Hyperparameter Analysis.}}
We study the impact of two critical hyperparameters: the number of group centroids per RQ‑KMeans stage ($k$) and the distillation‑loss weight ($\lambda$).

\begin{figure}[h!] 
    \centering
    \begin{subfigure}[b]{0.49\columnwidth}
        \centering
        \includegraphics[width=\textwidth]{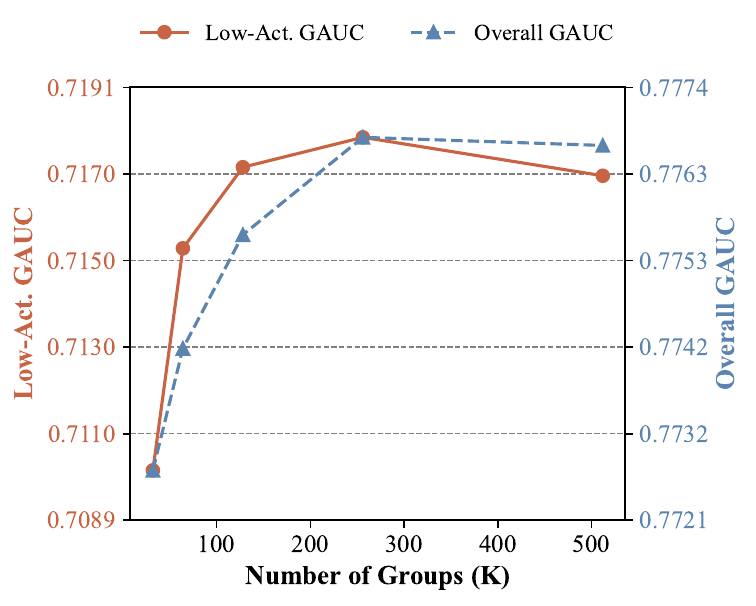} 
        \caption{Impact of Number of Groups}
        \label{fig:hyper_k}
    \end{subfigure}
    \hfill
    \begin{subfigure}[b]{0.49\columnwidth}
        \centering
        \includegraphics[width=\textwidth]{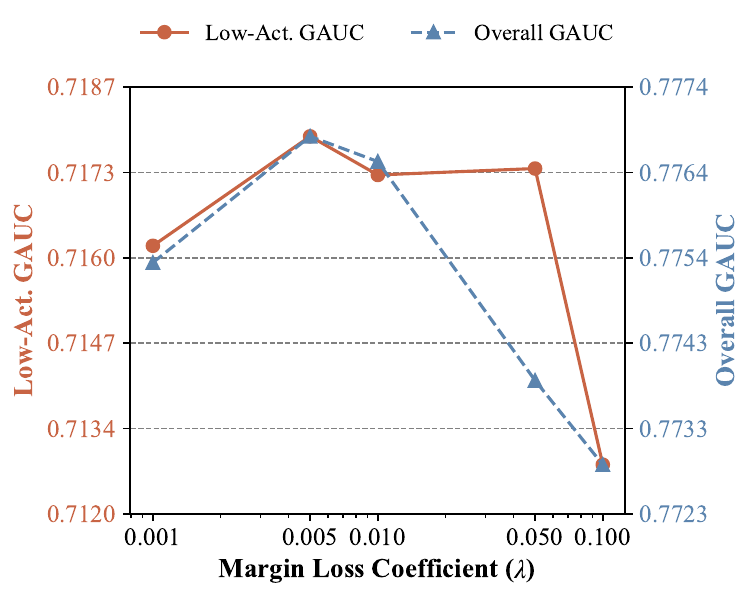} 
        \caption{Impact of Distillation Weight}
        \label{fig:hyper_lambda}
    \end{subfigure}
    
    \caption{Hyperparameter analysis for $k$ and $\lambda$.}
    \label{fig:hyperparameter_study}
\end{figure}

\begin{itemize}
    \item \textbf{Number of Group Centroids ($k$):} As shown in Figure~\ref{fig:hyper_k}, performance on both low‑activity and overall users improves as $k$ increases from 32 to 256, because larger $k$ produces finer‑grained semantic groups that capture more specific user interests. However, at $k=512$, performance drops because the resulting groups become overly sparse, reducing the stability and transferability of group‑level knowledge.
    
    \item \textbf{Distillation Loss Weight ($\lambda$):} Figure~\ref{fig:hyper_lambda} shows peak performance at $\lambda=0.005$. Smaller values weaken the regularization effect on the group channel, while larger values cause the distillation loss to interfere with primary CVR prediction, resulting in degraded overall performance.
\end{itemize}

\noindentparagraph{\textbf{Semantic Group Consistency.}}
To evaluate the reliability of the generated semantic groups, we conduct a quantitative consistency analysis. Results indicate a high degree of homogeneity: (i) basic user attributes within each group exhibit an average consistency rate of 83\%; and (ii) behavioral concentration is strong, with the top‑50 most purchased fine‑grained categories collectively accounting for an average of 60\% of all purchase behaviors in a group.

\noindentparagraph{\textbf{Adaptive Fusion Gate Behavior.}}
We analyze the operation of the adaptive fusion gate($\alpha_{\text{fusion}}$)  to confirm whether it behaves as intended. The average $\alpha_{\text{fusion}}$ for Level 1 (lowest‑activity) users is roughly 2.1× that for Level 5 (highest‑activity) users. This dynamic weighting indicates that the model relies more heavily on the group channel for data‑sparse users, validating the effectiveness of our adaptive balancing mechanism.

\subsection{Online A/B Testing}
\label{sec:online_test}

We evaluated \textit{ChoirRec} in a 21‑day live A/B test on the Taobao platform, following a two‑stage offline training pipeline. In the first stage, a data‑processing pipeline employed the LLM to construct semantic user groups and pre‑compute their aggregate priors. In the second stage, the \textit{ChoirRec} CVR model --- integrating hierarchical group‑ID fusion within its dual‑channel architecture --- was trained on these priors and then deployed for real‑time serving.

The results, shown in Table~\ref{tab:online_ab_test}, indicate clear business gains. Across all users, \textit{ChoirRec} delivered a +2.23\% increase in Orders and a +3.10\% increase in GMV. Improvements were most pronounced for low‑activity users, with Orders rising by +7.24\% and GMV by +9.27\%. Such uplifts in a production environment provide strong evidence that our approach effectively addresses user‑sparsity challenges while driving measurable gains in core business metrics.

\begin{table}[h]
\centering
\caption{Relative improvements (\%) in online metrics from a 21-day A/B test comparing ChoirRec with the Base Model.}
\label{tab:online_ab_test}
\begin{tabular*}{\columnwidth}{@{\extracolsep{\fill}} l ccc}
\toprule
\textbf{Online Metric} & \textbf{Low-Act.} & \textbf{High-Act.} & \textbf{Overall} \\
\midrule
Orders & +7.24\% & +1.56\% & +2.23\% \\
GMV & +9.27\% & +1.87\% & +3.10\% \\
Converting UV & +6.98\% & +1.15\% & +1.52\% \\
\bottomrule
\end{tabular*}
\end{table}












\section{Conclusion}

This paper addresses the challenge of conversion‑rate (CVR) prediction for low‑activity users by introducing \textit{ChoirRec}, an LLM‑driven semantic‑grouping framework. It leverages LLM to form high‑quality semantic user groups, creating a robust knowledge source to bridge the gap between high‑ and low‑activity users. A group‑aware hierarchical‑representation module generates rich group‑level priors to enrich sparse user embeddings, while a dual‑channel architecture ensures their effective integration into prediction.
Extensive large‑scale offline and online experiments show substantial gains for low‑activity users and consistent improvements across the entire user base, demonstrating that \textit{ChoirRec} is a practical and effective solution to user‑sparsity challenges in modern recommender systems. 


\bibliographystyle{ACM-Reference-Format}
\bibliography{sample-base}










\end{document}